\documentclass[twocolumn,10pt,prl,aps,showpacs,longbibliography]{revtex4-1}
\usepackage{graphicx}
\usepackage{amsbsy}
\usepackage[usenames, dvipsnames]{color}
\usepackage[utf8x]{inputenc}
\usepackage{amsmath,amssymb}

\def \LNj{{I^{(j)}_{\rm ncl}}}
\def \LNI{{I^{(1)}_{\rm ncl}}}
\def \LNII{{I^{(2)}_{\rm ncl}}}
\def \EI{{I_{\rm ent}}}
\def \GNI{{I_{\rm ncl}}}
\def \GNIT{{I_{\rm ncl}^{\rm tm}}}

\begin{document}
\title{Nonclassicality Invariant of General Two-Mode Gaussian States}
\author{Ievgen I. Arkhipov}
\email{ievgen.arkhipov01@upol.cz}
\address{RCPTM, Joint Laboratory of Optics of Palack\'y University and
Institute of Physics of CAS, Palack\'y University, 17. listopadu
12, 771 46 Olomouc, Czech Republic}
\author{Jan Pe\v{r}ina Jr.}
\address{RCPTM, Joint Laboratory of Optics of Palack\'y University and
Institute of Physics of CAS, Palack\'y University, 17. listopadu
12, 771 46 Olomouc, Czech Republic}
\author{Ji\v{r}\'i Svozil\'ik}
\address{RCPTM, Joint Laboratory of Optics of Palack\'y University and
Institute of Physics of CAS, Palack\'y University, 17. listopadu
12, 771 46 Olomouc, Czech Republic}
\author{Adam Miranowicz}
\address{Faculty of Physics, Adam Mickiewicz University, PL-61-614 Poznan, Poland}

\begin{abstract}
We introduce a new quantity for describing nonclassicality of an
arbitrary optical two-mode Gaussian state which remains invariant
under any global photon-number preserving unitary transformation
of the covariance matrix of the state. The invariant naturally
splits into an entanglement monotone and local-nonclassicality
quantifiers applied to the reduced states. This shows how
entanglement can be converted into local squeezing and vice versa.
Twin beams and their transformations at a beam splitter are
analyzed as an example providing squeezed light. An extension of
this approach to pure three-mode Gaussian states is given.
\end{abstract}
\maketitle

{\it Introduction.}---{ Despite of several decades of active
research, the nonclassical properties of light remain one of the
most intriguing problems in quantum optics} (for a review see,
e.g., Refs.~\cite{GlauberBook,Perina1994Book,MandelBook,
AgarwalBook}). A widely accepted criterion to distinguish
nonclassical states from the classical ones says that a quantum
state is nonclassical if its Glauber-Sudarshan $P$ function fails
to have the properties of a probability
density~\cite{Glauber63,Sudarshan63}.

For practical purposes, several operational criteria for
determining nonclassicality of either single-mode~\cite{
Miranowicz15,Shchukin05r,Richter02,Asboth05,Vidal02}  or
multimode~\cite{Miranowicz10,Bartkowiak11,Allevi2013,Filip13,Richter02,Vidal02}
fields have been derived using the fields'
moments~\cite{Vogel08,Shchukin05r,Miranowicz10,Allevi2013} or the
Bochner theorem \cite{Ryl2015}. Alternatively, the majorization
theory also provides useful criteria~\cite{Verma10}.
Nonclassicality can directly be identified according to its
definition when the quasidistributions of fields'
amplitudes~\cite{Lvovsky2009} or integrated
intensities~\cite{PerinaJr2013a} are reconstructed. The
nonclassicality, which can be revealed in the continuous variables
domain is becoming one of the most promising resourse for quantum
communication technologies~\cite{Weedbrook12}.

Up to now the two most widely studied kinds of nonclassical light
in the continuous variable domain are those exhibiting squeezing
and entanglement. Both kinds of light have recently been
recognized as potentially interesting not only for fundamental
physical experiments but also for many applications in quantum
technologies~\cite{NielsenBook,Horodecki09review,Polzik92,Weedbrook12,Wolfgramm10}.
Both squeezed and entangled light can easily be generated in
nonlinear processes, e.g., in second-subharmonic generation and
parametric down-conversion, respectively.

In these processes, the optical fields are generated in Gaussian
states. It has been shown in Refs.~\cite{Braunstein05,Plenio14}
that the Gaussian states obtained in both processes are mutually
connected by linear transformations easily accessible by `passive'
linear optics. A suitable linear transformation then allows to
obtain an entangled state at the expense of the original squeezed
state under suitable conditions. Also, entanglement can serve as
the source of squeezed light generated after suitable
linear-optical transformations. Here, we explicitly reveal the
conditions for the transformations of squeezed light into
entangled light and vice versa by constructing a suitable global
nonclassicality invariant (NI) that is composed of the additive
identifiers of entanglement and local nonclassicalities (e.g.
squeezing).

This allows rigorous control of the transformations of
nonclassical resources (encompassing both local nonclassicalities
and entanglement) in quantum-information protocols. Another
example of importance of our result is the capability of testing
the performance of schemes for the nonclassicality quantification
based on transforming local nonclassicalities into
entanglement~\cite{Asboth05}. Such schemes are considered as
important as the determination of, e.g., the Lee nonclassicality
depth~\cite{Lee91} or the Hillery nonclassical
distance~\cite{Hillery87}, which are commonly used as
nonclassicality measures, need the reconstruction of the
$P$~function. On the other hand, several measures of entanglement
are known both for discrete and continuous quantum
systems~\cite{Horodecki09review,Peres96,Horodecki97,Marian08,Eisert03,Adesso05,Vidal02}.
An intimate relation between entanglement and nonclassicality of,
in general, noisy twin beams has recently been revealed in
Ref.~\cite{arkhipov15}. A general approach for analyzing this
relation has been proposed in Ref.~\cite{Vogel14} considering
two-mode states. On the other hand, this NI allows to explicitly
determine the entanglement of a given Gaussian state through local
squeezing of the reduced single-mode states \cite{Adesso06}.

From the general point of view, entanglement implies global
nonclassicality of the overall field. On the other hand,
nonclassical multimode fields do not necessarily have to be
composed of mutually entangled parts. This occurs, when the parts
as such exhibit marginal (local) nonclassicalities. Examples
studied earlier have indicated that the action of global unitary
transformations may be viewed as a `certain flow' of entanglement
into local nonclassicalities and vice versa. We note that, in the
case of Gaussian fields, only the global unitary transformations,
which preserve the overall number of photons, are naturally
considered here. Such transformations are realized by passive
optical devices and, from the mathematical point of view, they
belong to the unitary group ${\bf U}(n)$. Indeed, there exists a
tight relation between entanglement and local nonclassicalities
which originates in the existence of a global nonclassicality
invariant which splits into entanglement and local
nonclassicalities quantifiers. In the past, an attempt to find
such NI for single-mode Gaussian states and the vacuum was done in
Ref.~\cite{Zubairy15} considering the logarithmic
negativity~\cite{Horodecki09review} as an entanglement measure and
the Lee nonclassicality depth as a local nonclassicality measure.
However, this approach worked only under quite specific
conditions. On the other hand, the approach based on a global
invariant succeeded when amplitude coherence and entanglement
quantified by the maximal violation of the Bell-CHSH inequality
have been analyzed together for a general two-qubit
state~\cite{Svozilik15}.

In this letter, considering two-mode Gaussian states, we reveal a
nonclassicality invariant resistant against any passive (i.e.,
photon-number preserving) unitary transformation of their
covariance matrix. We show that this invariant naturally
decomposes into the expressions giving the local nonclassicality
and entanglement quantifiers, which are monotones of the Lee
nonclassicality depth and the logarithmic negativity,
respectively. A global nonclassicality invariant is also suggested
and verified for pure three-mode states.

{\it Theory.}--- The characteristic function or, equivalently, the
corresponding complex covariance matrix $ {\bf A} $, can be used
for the description of a Gaussian bipartite state with its
statistical operator $\hat{\rho}$ as follows:
\begin{eqnarray}\label{CM}   
 {\bf A} = \begin{pmatrix}
  - B_{1} & C_1 & {\bar{D}}_{12}^{\ast}& D_{12} \\
  C_{1}^{\ast} & -B_1 & D_{12}^{\ast} & {\bar{D}}_{12} \\
  {\bar{D}}_{12} &D_{12} &-B_2& C_2 \\
  D_{12}^{\ast} &  {\bar{D}}_{12}^{\ast} & C_{2}^{\ast} & -B_2
 \end{pmatrix}. \\ \nonumber
\end{eqnarray}
The normally-ordered characteristic function is then expressed as
$ C_{\cal N}(\mbox{\boldmath$ \beta $})=
\exp\left(\mbox{\boldmath$ \beta $}^\dagger{\bf A}
\mbox{\boldmath$ \beta $}/2\right)$ using the vector $
\mbox{\boldmath$ \beta $} =
(\beta_1,\beta_1^*,\beta_2,\beta_2^*)^T$. Elements of the
covariance matrix $ {\bf A} $ in Eq.~(\ref{CM}) are defined as
\cite{Perina1991Book}
\begin{eqnarray} \label{qnc}   
 B_j&=&\langle\Delta\hat a^{\dagger}_{j}\Delta\hat a_{j}\rangle, \quad C_j=\langle\Delta\hat a_{j}^2\rangle,
  \quad j=1,2, \nonumber \\
 D_{12} &=&\langle\Delta\hat a_{1}\Delta\hat a_{2}\rangle, \quad \bar D_{12} =-\langle\Delta\hat
  a_{1}^{\dagger}\Delta\hat a_{2}\rangle
\end{eqnarray}
using the annihilation $(\hat a_j)$ and creation $ (\hat a_j^\dagger)
$ operators of mode $ j $, $ j=1,2 $.

The negative determinants $I_j = B_j^2-|C_j|^2$ ($ j=1,2 $) of the
diagonal blocks of the covariance matrix $ {\bf A} $ immediately
determine local nonclassicalities of modes 1 and 2. Indeed, the
Fourier transform of the normal characteristic function of mode 1
[2] given as $ C_{\cal N}(\beta_1,\beta_1^*,0,0) [$ $ C_{\cal
N}(0,0,\beta_2,\beta_2^*) $] diverges if $ I_1<0 $ [$ I_2<0 $].
Determinant $I_j$ is a monotone of the Lee nonclassicality depth
$\tau_j$ of mode $ j $ that is given as the maximal eigenvalue of
the $j$th diagonal block of the matrix $ {\bf A} $; i.e., $ \tau_j
= |C_j| - B_j$~\cite{Lee91}. Admitting also negative values for
$\tau_j$ which can quantify the distance from the
quantum-classical border we reveal the following monotonous
relation:
\begin{equation}\label{smn}  
 I_j = -\tau_j\left(\tau_j+2B_j\right).
\end{equation}
As the determinants $ I_j $ are invariant under local unitary
transformations, we may define the local nonclassicality
invariants (LNI) $\LNj=-I_j$, which quantify the local
nonclassicalities.

On the other hand, the separability criterion for a bipartite
state $\hat\rho$ derived in Ref.~\cite{Simon00,Marian01,Marian08}, which is based
on the positive partial transposition (PPT) of $\hat\rho$, can be
used to quantify the entanglement of $\hat\rho$ as
\begin{equation}\label{S}       
 I_{\rm ent}=I_{\mathcal S4}-\frac{1}{4}\tilde\Delta_{\mathcal S}+\frac{1}{16}\geq0,
\end{equation}
where $\tilde\Delta_{\mathcal S}=I_{\mathcal S1}+I_{\mathcal
S2}-2I_{\mathcal S3}$. Equality in Eq.~(\ref{S}) holds for
separable Gaussian fields. In Eq.~(\ref{S}), $I_{\mathcal S 1}$,
$I_{\mathcal S 2}$, and $I_{\mathcal S 3}$ are the local
invariants and $I_{\mathcal S 4}$ is a global invariant of the
covariance matrix $ {\bf A}_{\cal S} $ written for the symmetric
ordering of field operators. As shown below, the quantity $\EI$,
which we will call the entanglement invariant (EI), can serve as
an entanglement quantifier since it is a monotone of the
logarithmic negativity $ E_N $, i.e., it is also a monotone under
unitary transformations~\cite{Plenio05}.
 The invariants
$I_{{\cal S}k} $ of the symmetrically-ordered covariance matrix $
{\bf A}_{\cal S} \equiv
\begin{pmatrix} {\bf S}_1 & {\bf S}_{12} \\
 {\bf S}_{12}^{T}& {\bf S}_2 \end{pmatrix} $, as introduced in
Eq.~(\ref{S}), are determined as $I_{{\mathcal S}j}={\rm det}({\bf
S}_j)$, $ j=1,2 $, $I_{\mathcal S 3}={\rm det}({\bf S}_{12}) $, and
$ I_{\mathcal S 4}={\rm det}({\bf A}_{\cal S}) $.

The quantity $ \tilde \Delta_{\cal S}$
in Eq.~(\ref{S}), is related to the symplectic eigenvalue $ d_-
$ of the partially transposed covariance matrix $ {\bf A}_{\cal S}
$ as follows~\cite{Olivares12}
\begin{equation}\label{symd}    
 d_- = \frac{1}{\sqrt{2}}\sqrt{\tilde\Delta_{\cal S}-\sqrt{\tilde\Delta_{\cal S}^2-4I_{\mathcal S4}}}.
\end{equation}
Combining Eqs.~(\ref{S}) and (\ref{symd}) we arrive at
\begin{equation}\label{symdf}   
 d_- =  \frac{1}{\sqrt{2}}\sqrt{I'-\sqrt{I'^2-4I_{\mathcal S 4}}},
\end{equation}
where $I'=4I_{\mathcal S 4}+4\EI+1/4$. The eigenvalue $ d_- $ then
gives the logarithmic negativity $E_N$ as follows
\begin{equation}\label{logneg}  
 E_N={\rm {max}}[ 0,- \ln(2d_-) ].
\end{equation}
For pure states, we have $I_{{\cal S} 4}=1/16$ and the following
monotonous relation between logarithmic negativity $ E_N $ and
entanglement invariant $ \EI $ can be given:
\begin{equation}\label{NvsS}  
 E_N={\rm {max}}\left[0,{\rm {ln}}\left(2\sqrt{\EI}+\sqrt{1+4\EI}\right)\right].
\end{equation}
A detailed analysis of Eq.~(\ref{symdf}) confirms that, by keeping
the global invariant $ I_{{\cal S} 4} $ fixed, the EI $\EI$
remains a monotone of the logarithmic negativity $E_N$ even for
general two-mode Gausssian states.

It is easy to show that the global nonclassicality invariant (GNI)
$\GNI$ defined as
\begin{equation}\label{law}     
 \GNI= \LNI +\LNII + 2\EI
\end{equation}
is invariant under any global passive unitary transformation
applied simultaneously to both covariance matrices $ {\bf A} $ and
$ {\bf A}_{\cal S} $. Using the definitions of $\LNI$, $\LNII$,
and $\EI$, together with the fact that the local invariant $
I_{{\cal S} 3} $ does not depend on operator ordering, we have
\begin{eqnarray}                
 \GNI &=& -I_1-I_2-2I_{\mathcal S4}+\frac{1}{2}(I_{\mathcal S1}+I_{\mathcal S2}-2I_{\mathcal S3})-\frac{1}{8} \nonumber \\
     &=& -\Delta +\frac{1}{2}\Delta_{\cal S}-2I_{\mathcal S
  4}-\frac{1}{8}.
\label{10}
\end{eqnarray}
In Eq.~(\ref{10}), $ \Delta_{\cal S} =I_{{\cal S}1}+I_{{\cal S}2}
+ 2I_{\mathcal S3}$ represents the global invariant of the
symmetrically-ordered covariance matrix, whereas the quantity $
\Delta =I_1+I_2+2I_{\mathcal S3}$ gives the global invariant of
the normally-ordered covariance matrix.

For pure two-mode Gaussian states we have $
\Delta_{\cal S} = 1/2$, $I_{{\cal S} 4}=1/16$, $ \GNI =
-\Delta =B_1+B_2$, and $ \EI = - I_{{\cal S} 3} $. Therefore in this case,
the GNI $\GNI$ is determined by invariants of the normally-ordered CM.

We note, that our invariant can also be applied to a single-mode
Gaussian state. Specifically, this is a special case of our two-mode
analysis if we assume that one of the input modes to the beam
splitter (shown in Fig.~1) is in the vacuum state. This case is in 
analogy to the original approach of Asboth et al.~\cite{Asboth05}.

According to Eq.~(\ref{law}), which gives the central result of
this paper, any passive unitary transformation modifies in general
the LNIs $\LNI$ and $\LNII$ as well as the EI $\EI$, such that the
value of the GNI $\GNI$ is unchanged. During such a
transformation, the decrease (increase) of the local
nonclassicalities has to be compensated by the increase (decrease)
of entanglement. Thus, formula~(\ref{law}) represents a
conservation law of the nonclassicality.


{\it Example: A twin beam (TWB) at a beam splitter.} --- TWBs are
provided by parametric down-conversion and, in their noiseless
variant, are composed of many photon pairs with the twin
photons embedded in the signal and idler fields. This guarantees
strong entanglement in a TWB. As the marginal fields are thermal,
no local nonclassicality is observed. Mixing of the signal and
idler fields at the beam splitter represents a unitary
transformation that modifies both entanglement and local
nonclassicality as follows (for the setup, see Fig.~1).
\begin{figure} [t!]     
\includegraphics[width=0.48\textwidth]{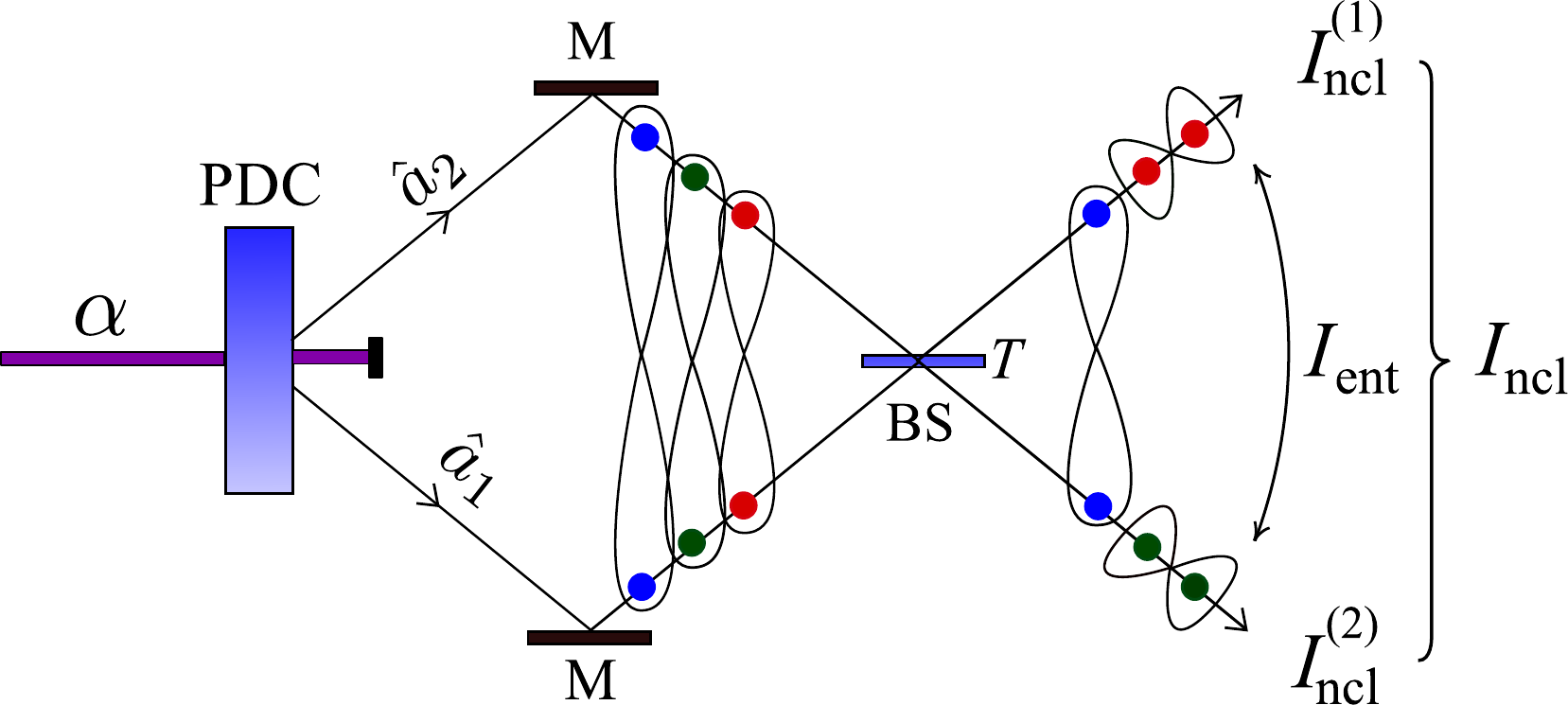}
 \caption{(Color online). Pump field $ \alpha $ generates photon pairs in the signal ($\hat a_{1}$) and
 idler ($\hat a_2$) fields via parametric down-conversion (PDC). Photon pairs are mixed on a beam splitter (BS) with transmissivity $T$:
 photons in a pair either stick together (bunch) to contribute to squeezing or remain in different beam-splitter ports (antibunch) to form
 entanglement.}
\end{figure}

The LNIs $ \LNj $ and EI $\EI$ acquire the form
\begin{eqnarray}\label{VSpure}  
 \LNj &=& - B^2_{\rm p}+4T(1-T)(B^2_{\rm p}+B_{\rm p}), \hspace{3mm} j=1,2, \nonumber \\
 \EI&=& (2T-1)^2(B^2_{\rm p}+B_{\rm p}),
\end{eqnarray}
where $B_{\mathrm p}$ is the mean photon-pair number. According to
Eq.~(\ref{VSpure}), the LNIs $ \LNj $ are given by two terms. The
first (negative) term arises from the input thermal statistics and
describes photon bunching. The second (positive) term is much more
interesting as it describes the squeezing effect at a
beam-splitter output port. At the `microscopic level', this effect
originates in pairing of photons in the output port caused by
sticking of two twin photons at the beam
splitter~\cite{MandelBook,Paris97,Braunstein05}. Such local
pairing of photons creates local nonclassicalities of the field.
The `sticking effect' at the beam splitter reduces the number of
photon pairs with photons found in different output ports and, so,
it naturally reduces their entanglement, in agreement with
Eq.~(\ref{VSpure}). The strength of the relation between the
micro- and macroscopic pictures is revealed when the formula for
the GNI in Eq.~(\ref{law}) is written, $ \GNI = 2B_{\rm p} $. The
GNI being linearly proportional to the number of photon pairs
clearly shows that, in case of TWBs, only individual photon pairs
are responsible for their entanglement and local
nonclassicalities.

Analyzing Eq.~(\ref{VSpure}), the maxima in the LNIs $ \LNj $ are
reached for the balanced beam splitter ($ T=1/2 $) that does not
allow any entanglement~\cite{Paris97}. The more unbalanced is the
beam splitter, the greater is the $ \EI $ and also the smaller are
the LNIs $ \LNj $. Local nonclassicalities of the output fields
occur only for $ |T-1/2| < 1/(2\sqrt{B_{\rm p}+1}) $. The
quantification of this behavior is done in the graphs of Fig.~2
showing the LNIs $ \LNj $ and EI $ \EI $ as functions of the mean
photon-pair number $ B_{\rm p} $ and transmissivity $ T $.
\begin{figure} [t!]     
\includegraphics[width=0.4\textwidth]{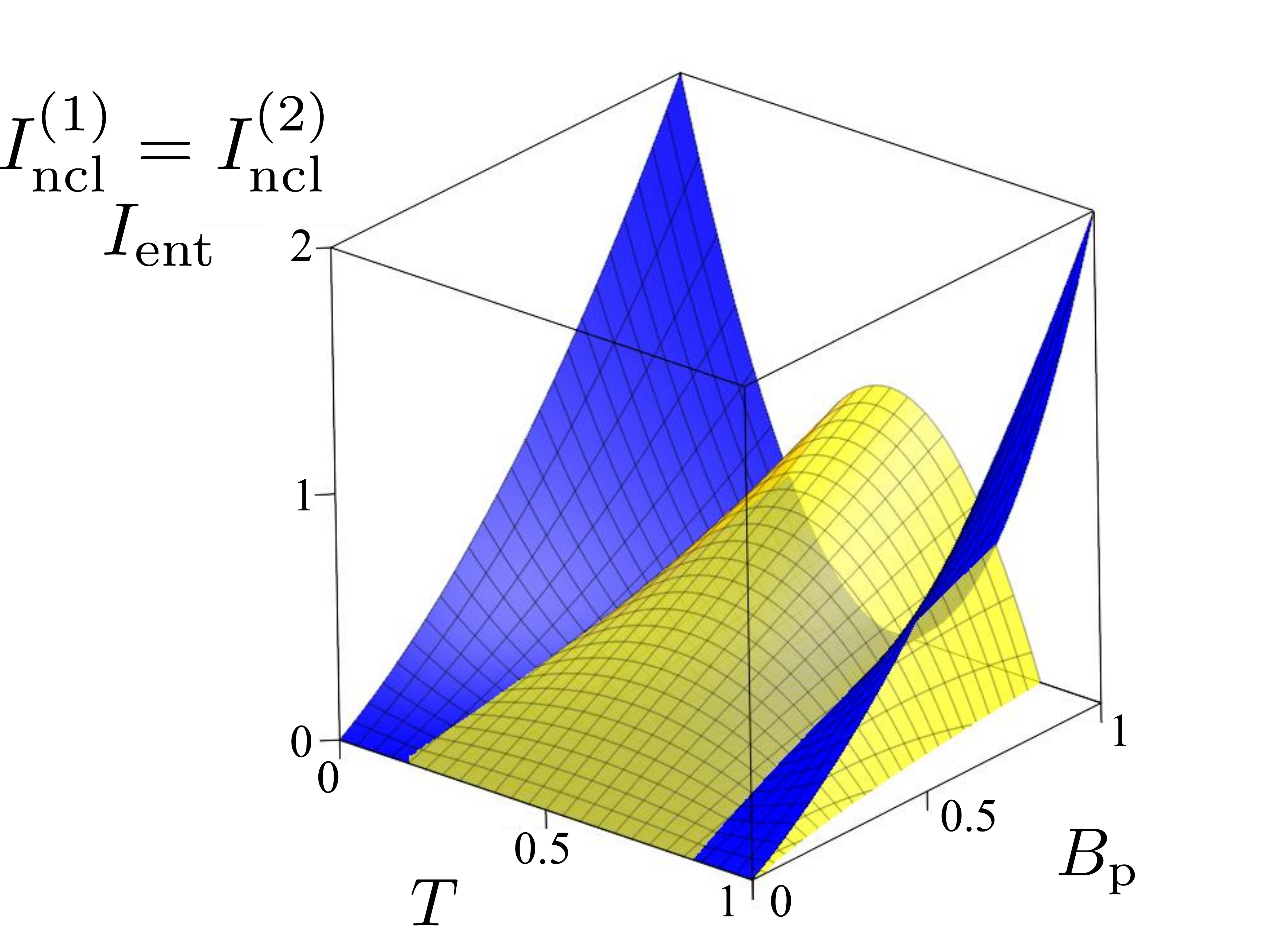}
 \caption{(Color online). Local nonclassicality invariants $ \LNI=\LNII $
 [yellow (light) surface] and entanglement invariant $ \EI $ [blue (dark) surface] as
 functions of the mean photon-pair number $B_{\rm p}$ and transmissivity $T$ for twin beams (only positive values are plotted).}
\end{figure}

We note that, similarly as the input TWB may provide squeezed
light at the beam-splitter outputs, the incident squeezed light
present in one or both input ports allows for the generation of
the entangled output fields.


{\it Extension to pure three-mode Gaussian states.}--- Motivated
by the results for two-mode Gaussian states, we suggest an
appropriate form of a three-mode NI relying only on the LNIs and
pairwise (two-mode) EIs. The proposed NI is invariant under any
global passive unitary transformation provided that only pure
three-mode Gaussian states are considered. This observation
accords with the results in
Refs.~\cite{Adesso06,Serafini05,Braunstein05} showing that (a) any
entangled three-mode state can be transformed via a global unitary
transformation into a state of three independent squeezed modes
and (b) genuine three-mode entanglement can be expressed through
the two-mode entanglements of three subsystems obtained by the
reduction with respect to one mode. We note that this result
applies also to the symmetric GHZ state in the continuous domain.

The symmetrically-ordered covariance matrix $ {\bf A}_{\cal
S}^{(3)} $ of a three-mode Gaussian state is written as
\begin{equation}      
 {\bf A}_{\cal S}^{(3)}= \begin{pmatrix}
  {\bf S}_{1} & {\bf S}_{12} & {\bf S}_{13} \\
  {\bf S}_{12}^T & {\bf S}_{2} & {\bf S}_{23} \\
  {\bf S}_{13}^T & {\bf S}_{23}^T  & {\bf S}_{3}
  \end{pmatrix},
\end{equation}
where the matrix $ {\bf S}_j $ describes mode $ j $ and matrix $
{\bf S}_{jk} $ characterizes the correlation between modes $ j $
and $ k $. The matrices $ {\bf S}_{jk} $ are independent of the
operator ordering and, so, they occur also in the normally-ordered
covariance matrix $ {\bf A}^{(3)} $. We construct the three-mode
GNI $\GNIT $ as follows
\begin{equation}\label{law_3m} 
 \GNIT = \sum_{j=1}^{3}\LNj + 2 \sum_{k>j=1}^{3} I_{\rm ent}^{(jk)},
\end{equation}
where $\LNj$ is the LNI of mode $ j $ and $ I_{\rm ent}^{(jk)}$ is
the EI of modes $ j $ and $ k $ determined from their reduces
statistical operator. Equation~(\ref{law_3m}) can be rewritten as $
\GNIT = -\Delta^{(3)} + \Delta^{(3)}_{\cal S}/2 - K - 3/8 $, where
$ \Delta^{(3)} = -\sum_{i=1}^{3}\LNj + 2\sum_{i<j,1}^{3}{\rm
det}({\bf S}_{ij}) $, $ \Delta^{(3)}_{\cal S} = \sum_{i=1}^{3}
{\rm det}({\bf S}_i) + 2 \sum_{i<j,1}^{3}{\rm det} ({\bf S}_{ij})
$, and $ K = 2 \sum_{i<j,1}^{3}{\rm det}({\bf A}_{{\cal
S}ij}^{(2)}) - \sum_{k=1}^{3}{\rm det}({\bf S}_k)/2 $ with $ {\bf
A}_{{\cal S}ij}^{(2)} = \begin{pmatrix} {\bf S}_{j} & {\bf S}_{jk}
\\ {\bf S}_{jk}^T & {\bf S}_{k} \end{pmatrix} $.
Since $ \Delta_{\cal S}^{(3)}=3/4$ and $ {\rm det}({\bf A}_{{\cal
S}ij}^{(2)}) = {\rm det}({\bf S}_k)/4 $ for pure three-mode
states, we have $\GNIT = - \Delta^{(3)} $. As $ \Delta^{(3)} $ is
a global invariant of the normally-ordered covariance matrix $
{\bf A}^{(3)} $ under passive unitary transformations, the GNI
$\GNIT $ becomes unchanged when such transformations are applied.
Similarly as for pure two-mode states, we have $\GNIT =
\sum_{i=1}^{3} B_{i}$, where $B_i$ gives the mean number of
photons in mode $ i $. Therefore the GNI for pure three-mode state
is determined by the local invariants of the normally-ordered
covariance matrix $ {\bf A}^{(3)} $. Formula~(\ref{law_3m}) for
the pure three-mode GNI $ \GNIT $ shows that the three-mode
entanglement can be quantified by the sum of three two-mode
entanglements. Monitoring the three LNIs and three EIs involved in
Eq.~(\ref{law_3m}) allows to quantitatively analyze the evolution
of nonclassicality resources in any quantum-information protocol
described by passive unitary transformations.

We note that the generalization to the case of $ m>3 $ modes based
on the assumption of two-mode entanglement quantifiers
($K=2\sum\limits_{i<j}^{m}{\bf
S}_{ij}-\frac{m-2}{2}\sum\limits_{k=1}^{m}{\bf S}_{k}$) is not
useful since the obtained quantity is not a global invariant,
similarly as in the case of mixed three-mode states.


{\it Example: A twin beam transformed by two beam splitters}--- A
simple method providing varying bipartite entanglement among three
output ports as well as locally nonclassical output fields can
easily be constructed from the previous example of a TWB at a beam
splitter. We enrich this method by additional splitting the field
at the output port 2 by a balanced beam splitter with the output
ports 2 and 3 (for the scheme, see Fig.~3)
\cite{Braunstein1998,Loock2000}.
\begin{figure} [t!]     
\includegraphics[width=0.45\textwidth]{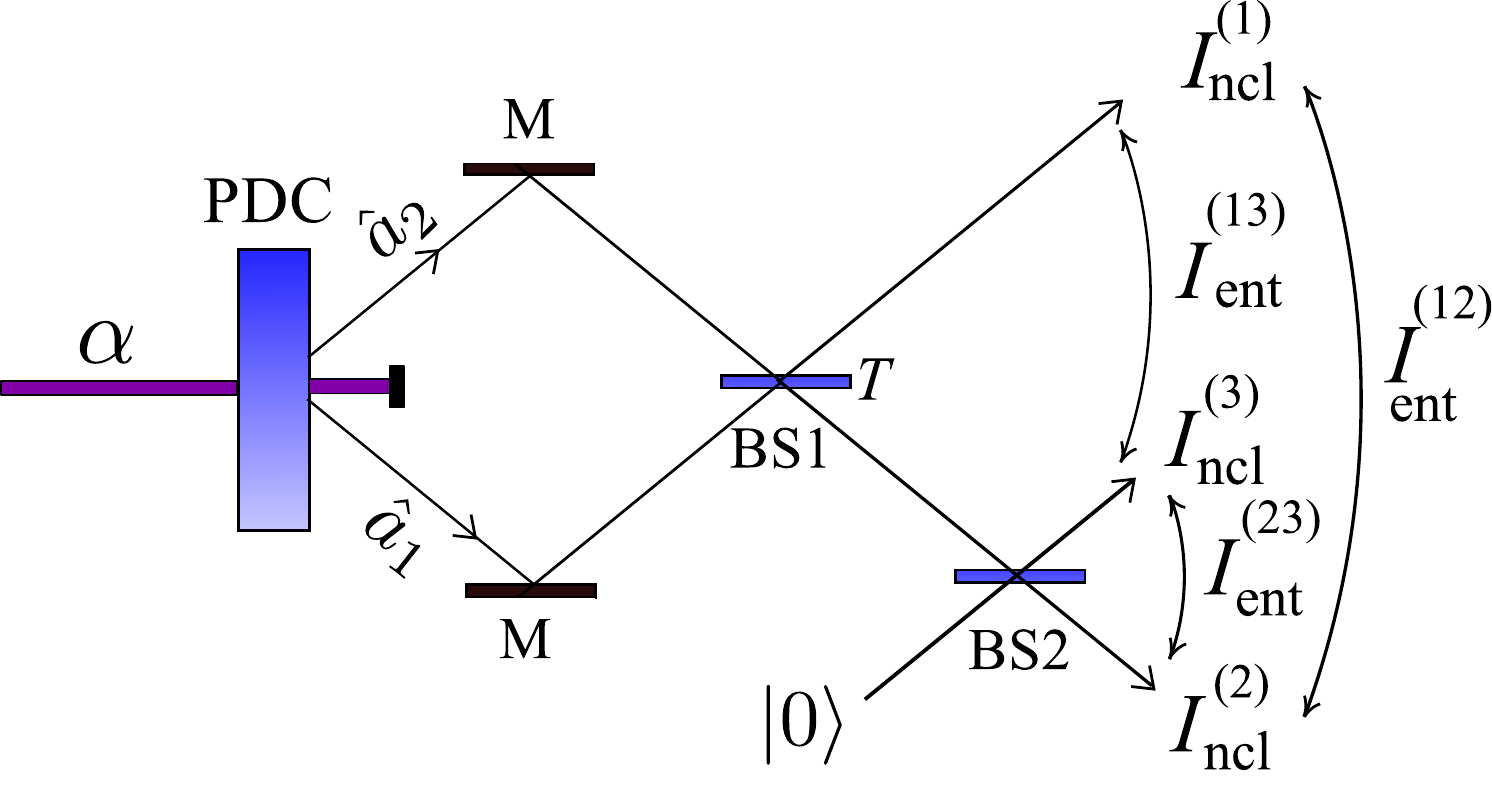}
 \caption{(Color online). Pump field $ \alpha $ generates photon pairs in the signal ($\hat a_{1}$) and
 idler ($\hat a_2$) fields via parametric down-conversion (PDC). Photon pairs are mixed on a beam splitter (BS) with transmissivity
 $T$. Field in one output port of this beam splitter is combined
 with the vacuum $ |0\rangle $ at another balanced beam splitter.
 LNIs $ I_{\rm ncl}^{(j)} $ and EIs $ I_{\rm ent}{(jk)} $ characterized the three output fields.}
\end{figure}
This results in a general three-mode state. From the point of view
of entanglement, photon pairs, which are originally responsible
for the entanglement between modes 1 and 2, are divided by the
second beam splitter to those establishing entanglement either in
modes 1 and 2, or modes 1 and 3. On the other hand, the photon
pairs, which are localized in mode 2 and responsible for its
squeezing, may split at the second beam splitter giving rise to
the entanglement between modes 2 and 3. This results in a full
three-mode entanglement. Indeed, the presented theory provides the
following formulas:
\begin{eqnarray}\label{VSpure2}  
 & I^{(1)}_{\rm ncl} = \frac{1}{4} \LNj = \frac{1}{4} I^{(23)}_{\rm ent} =
  - B^2_{\rm p}+4T(1-T)(B^2_{\rm p}+B_{\rm p}),&  \nonumber \\
 & I^{(1j)}_{\rm ent} = \bigl[\frac{1}{2} -2T(1-T)\bigr] (B^2_{\rm p}+B_{\rm p}),
  \hspace{3mm} j=2,3. &
\end{eqnarray}
These formulas are visualized in  Fig.~3, which confirm our
predictions. For the transmissivities $ T $ in certain interval
found in the previous example and excluding $ T=1/2 $, we have a
genuine three-mode entanglement. Moreover all the three output
fields are locally nonclassical. Whereas the LNIs $ I^{(j)}_{\rm
ncl} $ decrease with the increasing unbalance of the first beam
splitter, the decrease of the EI $ I^{(23)}_{\rm ent} $ is
compensated by the increase of the EIs $ I^{(12)}_{\rm ent} $ and
$ I^{(13)}_{\rm ent} $. We note that the GNI is again linearly
proportional to the initial photon-pair number $ B_{\rm p} $,
$\GNIT = 2B_{\mathrm p}$.

{\it Critical analysis of the Asboth et al. scheme for
nonclassicality quantification}--- If $ T=1/2 $ in the above
example, two separable squeezed states beyond the first beam
splitter occur and, so, we retain the standard Asboth {\it{et
al.}} approach~\cite{Asboth05} for the nonclassicality
quantification for the field in output port 2 of the first beam
splitter. As certain amount of squeezed photon pairs remains in
the output fields 2 and 3 beyond the second beam splitter, the
standard approach cannot provide a full quantification of the
nonclassicality of the analyzed field. Nevertheless, the EI $
I^{(23)}_{\rm ent} $ accessible in the Asboth {\it {et al.}}
method provides a good estimate of the nonclassicality of the
analyzed field since, according to Eq.~(\ref{VSpure2}), the LNI $
\tilde I^{(2)}_{\rm ncl} \equiv I^{(2)}_{\rm ncl} + I^{(3)}_{\rm
ncl} + 2 I^{(23)}_{\rm ent} $ is linearly proportional to the EI $
I^{(23)}_{\rm ent} $ for an arbitrary transmissivity $ T $.

\begin{figure}     
 \includegraphics[width=0.23\textwidth,keepaspectratio=true]{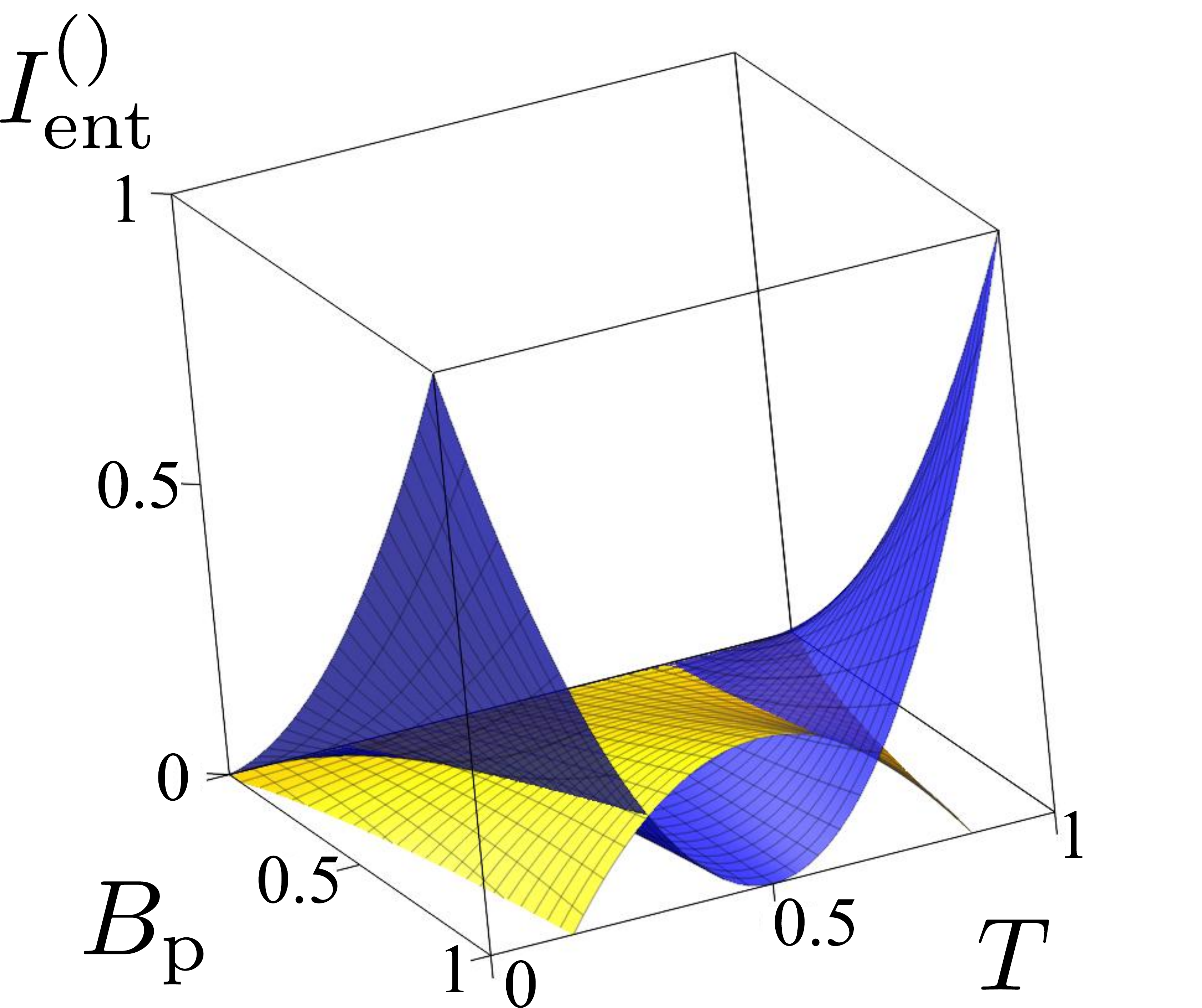}
 \includegraphics[width=0.23\textwidth,keepaspectratio=true]{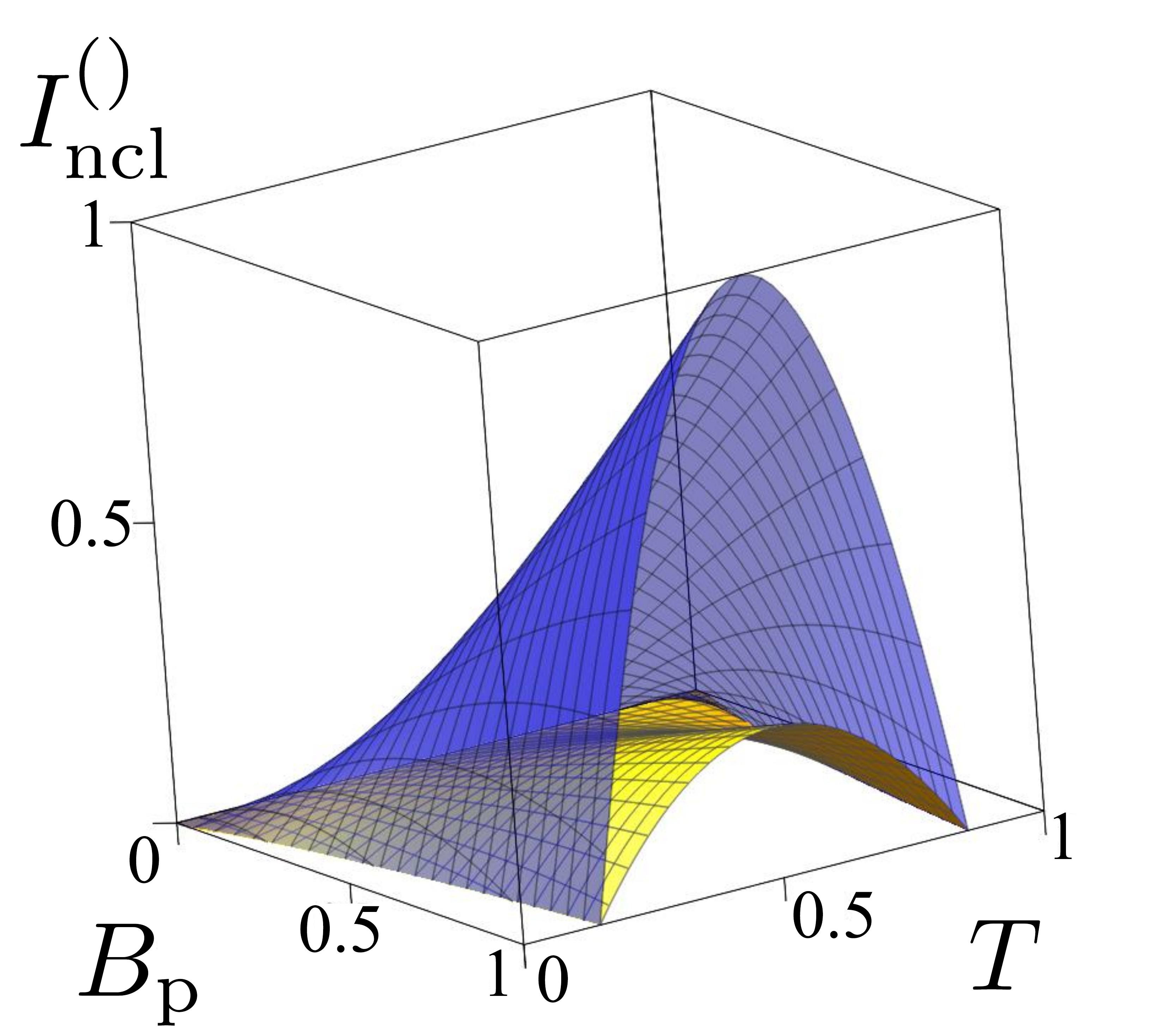}
 \centerline{ (a) \hspace{3cm} (b) }

 \caption{(Color online). (a) Entanglement invariants $ I^{(12)}_{\rm ent}=I^{(13)}_{\rm ent}$ [blue dark
 surface] and $ I^{(23)}_{\rm ent}$ [yellow light surface]
 and (b) local nonclassicality invariants $\LNI$ [blue dark surface] and $ \LNII=I^{(3)}_{\rm ncl}$ [yellow light surface]
 as they depend on the mean photon-pair number $B_{\rm p}$ and the beam-splitter transmissivity $T$
 for an initial pure TWB in the scheme of Fig.~3 (only positive values are shown).}
\end{figure}

{\it Conclusion.}--- We have found an invariant for general
two-mode Gaussian states which comprises the terms describing both
marginal nonclassicalities of the reduced states and the
entanglement of the whole system. Those terms being monotones
under any unitary transformation of the Lee nonclassicality depth
and the logarithmic negativity, respectively, quantify the flow of
nonclasical resources when passive unitary transformations are
applied. We gave the extension of these results to pure three-mode
Gaussian states. As examples, we found a relation between twin
beams and squeezed states. Moreover we critically analyzed the
Asboth {\it {et al.}} method for quantifying nonclassicality.

{\it Acknowledgments.}--- The authors thank Jan Pe\v{r}ina and Anirban
Pathak for discussions. They acknowledge support from the projects
15-08971S of GA \v{C}R and LO1305 of M\v{S}MT \v{C}R. I.A. thanks
project IGA\_PrF\_2015\_004 of IGA UP Olomouc.

\noindent {\bf Author contributions statement} All authors
contributed to the development of the theory as well as the
preparation of the manuscript.

\noindent {\bf Additional information}

\textbf{Accession codes}; \textbf{Competing financial interests}
The authors have no competing financial interests.

%

\end{document}